# Immersive Anatomical Scenes that Enable Multiple Users to Occupy the Same Virtual Space: A Tool for Surgical Planning and Education

Alex J. Deakyne[1,2], Erik N. Gaasedelen[1,2], Tinen L. Iles[2], Paul A. Iaizzo[2,3]

Departments of [1]Bioinformatics and Computational Biology, [2]Surgery, and [3]Biomedical Engineering, University of Minnesota

## Abstract

3D modeling is becoming a well-developed field of medicine, but its applicability can be limited due to the lack of software allowing for easy utilizations of generated 3D visualizations. By leveraging recent advances in virtual reality, we can rapidly create immersive anatomical scenes as well as allow multiple users to occupy the same virtual space: i.e., over a local or distributed network. This setup is ideal for pre-surgical planning and education, allowing users to identify and study structures of interest. I demonstrate here such a pipeline on a broad spectrum of anatomical models and discuss its applicability to the medical field and its future prospects.

## Introduction

As the numbers of distributed internet technologies grow, innovation increasingly aims to solve the problem of allocation of resources. Depending on the application, such technologies can be used to serve more users by reducing the effective distances between the distributor and customer. The global physician population, who's specialization become more narrow and complex, is one such example where such technological applications has been under-optimized. Yet, proper medical care is frequently reliant on primary physician, consultants, specialized imaging equipment, and a given patient being present in the same place. This naturally leads to the question of how to decouple these spatial dependencies while maintaining or even enhancing the 'quality of care'.

To date, companies participating in the telemedicine market have primarily focused on developing smartphone applications which have allowed the physician-patient interview to take place over large distances. These applications can increase their analytical capabilities by acquiring data from sensing devices such as stethoscope, EKG, or EEG. For fields like interventional cardiology or surgery, a fully remote interface has yet to become applicable for tasks which require high human dexterity. However, physician collaborations has been improved through technologies like Google Glass, which can stream a live clinical procedures to a remotely consulting physician. But typically to date, these kinds of interactions force the remote physician to have a narrower perceptual awareness of the procedure, utilizing video with a small field of view and, presumably, a written preclinical plan. Consequently, greater technological utilizations are not yet commonplace.

Today advancement in medical imaging have been coupled with the field of radiology to successfully allow for the practical separation of patient diagnoses into remote observations and distributed interpretations. Yet for example, when such imaging approaches are so to be used for surgical planning, the utility of remotely consulting physicians can become diminished. We propose that this is because descriptions concerning spatial awareness requires a greater communication bandwidth between physician and consultant, placing a cognitive strain that prevents widespread adoption. This could be much easier, however, if there were modalities that could supplement spoken communication with rich spatial awareness and gestural abilities.

The fields of virtual and augmented realities are rapidly maturing even within the timeframe of months, driven primarily by the video gaming market. A wealth of tools dedicated to creating effective virtual reality environments has followed, even allowing developers to incorporate internet-based multiplayer features into these scenes.

Virtual and augmented reality systems have differences in how they display information to the user. Virtual reality utilizes a headset which completely occludes the user's native field of view, replacing it with high resolution screens that create an artificial space, based on the headsets spatial coordinates. Augmented reality, in contrast, allows the user to maintain their environmental awareness of the real world, by instead projecting holograms in their native space. Depending on the applications and constraints set by AR/VR systems, each can have their own advantages and disadvantages.

Meanwhile, the field of medical 3D modeling continues to boom, aiming to allow physicians to have a more intimate understanding of their patients' relational anatomies. This creates a natural progression to combine the advancements of virtual reality and anatomical 3D modeling, into a single pipeline. Researchers have already considered VR for surgical planning (Robinoy et al, 2007). However, its adoption has been limited in part because there are not many tools yet available to take advantage of this modality. However, by combining networking functionalities, 3D clinical models, physical prints, and virtual reality, we can fulfill many of the requirements set above for an effective distributed surgical preplanning and consultation tool: a mixed reality approach. We propose that developments of such technologies will not only allow for more effective communication between physicians, but will provide a substrate for novel collaborative pipelines in both patient care and education. (Figure 1)

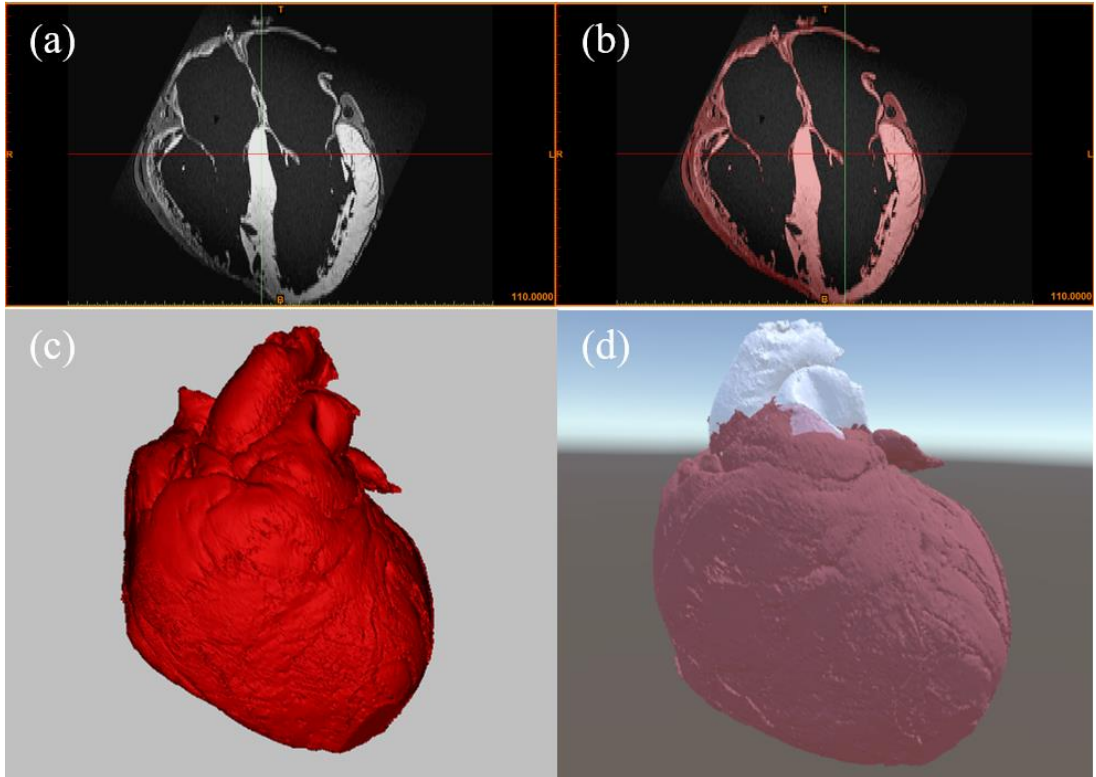

**Figure 1: Flowchart diagram of proposed mixed reality methodology. (a) A DICOM scan in MIMICS depicting a cross sectional view of a heart. (b) 2D mask of the heart tissue generated through thresholding functions. (c) The resulting 3D model of the heart created by the 2D masks which can then be 3D printed. (d) The same 3D model of the heart is imported into Unity and a virtual reality environment of this given heart is generated.**

# Methods

### *3D Model Development*

Voxel-based medical imaging can be retrieved from any compatible CT or MRI scanner: these are required to create corresponding 3D models of the patient's anatomy. Yet, these differing modalities can affect the resultant quality of the final model because of differing contrasts and resolutions. In our research group, we can scan static hearts that have been explanted using MRI. We can also visualize scans from cadaveric sources in which we can inject contrast to better model the vasculature. This does not affect the technical contributions of this report, but shows the diversity for which this process may be applicable.

Segmentation of various obtained medical imaging was performed using the Mimics (Materialise, Leuven Belgium). This process is characterized by the creation of masks which either represent the complete model or specified anatomical features. The creation of these masks typically requires exceptional care, as those who are not experienced with medical imaging and anatomical identification may produce different results. Additionally, it is also required that anatomical structures which must be either manipulated or colored uniquely within the virtual reality environment, thus must be segmented with its own mask. 3D models can be generated from these masks, but to achieve realistic models, some post processing is typically performed. This commonly involves filling of holes, or the smoothing of rough edges. The validation of such operations and their effects on anatomical realism is an active field of investigation and advances will provide immediate improvements to our pipeline.

### *Required Hardware and Software*

To allow these models to be intractable in virtual reality, the Unity3D (Unity Technologies, San Francisco, CA, USA) video game engine was used along with the SteamVR API (Valve Corp., Bellevue, WA, USA) for the HTC Vive . Yet, neither Unity3D nor the HTC Vive are solely required for the creation of such virtual reality scenes. For example, the Unreal Engine (Epic Games, Cary, NC, USA) and the Oculus Rift (Oculus VR, LLC, Irvine, CA, USA) are both competing technologies which offer as a comparable alternative. Nevertheless, interactive components required the creation of scripts which allowed the user to change their coordinate position or navigate to a different virtual environment. For our setup, a computer with a capable graphics card and CPU were necessary to render objects which were intractable in real time. In our case we used an Nvidia GTX-1070 graphics card with Intel core-i7 7700k CPU.

### *Development of Multiplayer Functionality*

Multiplayer functionality required the specific programming of network features. For this we used the Photon Bolt (https://www.photonengine.com/bolt) networking engine which provides simple abstractions for monitoring remote states. To properly set this up within a virtual reality scene, multiple steps had to be taken. First, an avatar was created that would represent each player's headset and controller to every client connected to the server. In this case we simply used a circle and rectangular prism to represent a face with a visor, and a rectangular prism with laser to represent the controller (Figure 2).

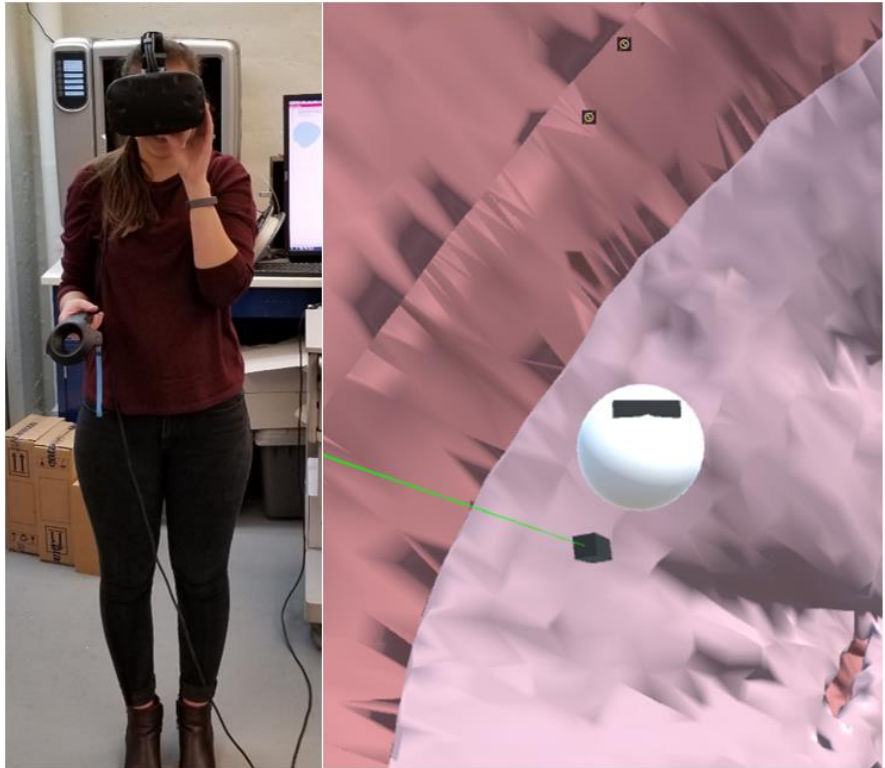

**Figure 2: Image of a user wearing VR headset next to their virtual avatar.**

A custom script was created that spawns an avatar on the server whenever a user enters a new scene. This script utilizes hooks for the headset and controller that allows them to be called when they first appear in a scene. Once they can be referenced, an avatar is instantiated on the server at the user's current position and is replicated to all other connected VR users/clients. This script also sets the avatar transform to that of the user, ensuring the transform information of the avatar is always the same as the user's headset and controller.

Next, a network state must be created for our avatar to define its network properties. This can be easily setup through Bolt. In this project, we needed to further define the transform as a property of our avatar's state. We set the transform to be replicated to everyone so that a user's avatar is replicated to every client connected to the server (Figure 3a). Lastly, we add a Bolt Entity component to the avatar and set it equal to the transform state that we had previously defined. Now, once the avatar is spawned on the network at the user's location, any local movement done by the user is replicated by their avatar to all clients connected to the server (Figure 3b).

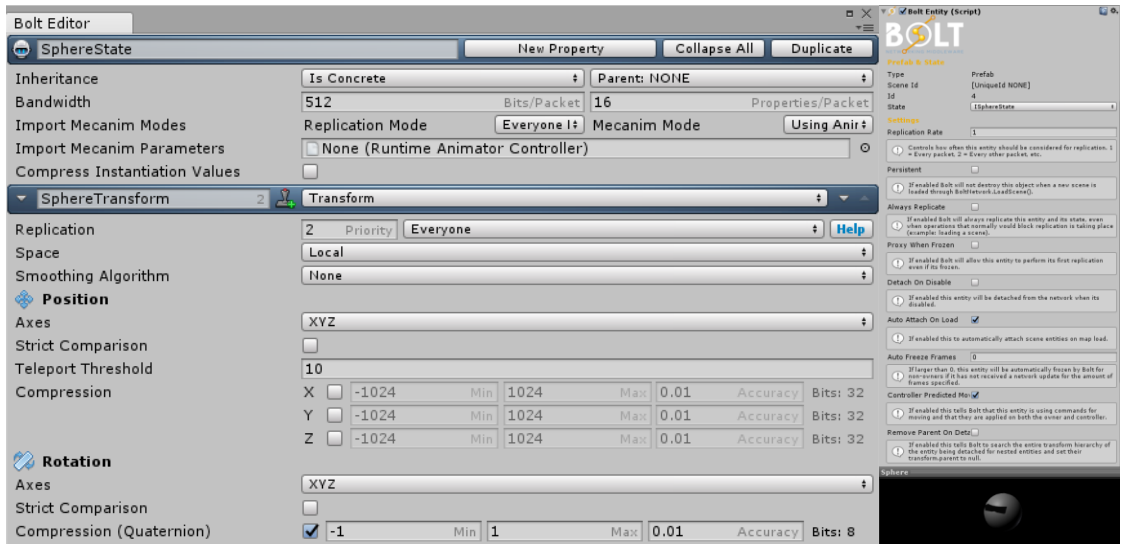

<div align="center">(a)                              (b)</div>

**Figure 3: (a) The network state created for our avatar. This state contains the sphere transform property we defined. (b) A bolt entity was added to our avatar, defining its network state.**

### *User Interface*

A VR scene was created that displays a Menu. This user interface (UI) allows users to select different virtual scenes to explore (Figure 4). A script was created that utilizes the laser pointer on the controller to select a scene to enter. The laser is a ray-cast object extending from the tip of the HTC Vive controller and the buttons on the menu detect collision with the laser. When a collision occurs, the button becomes opaque.

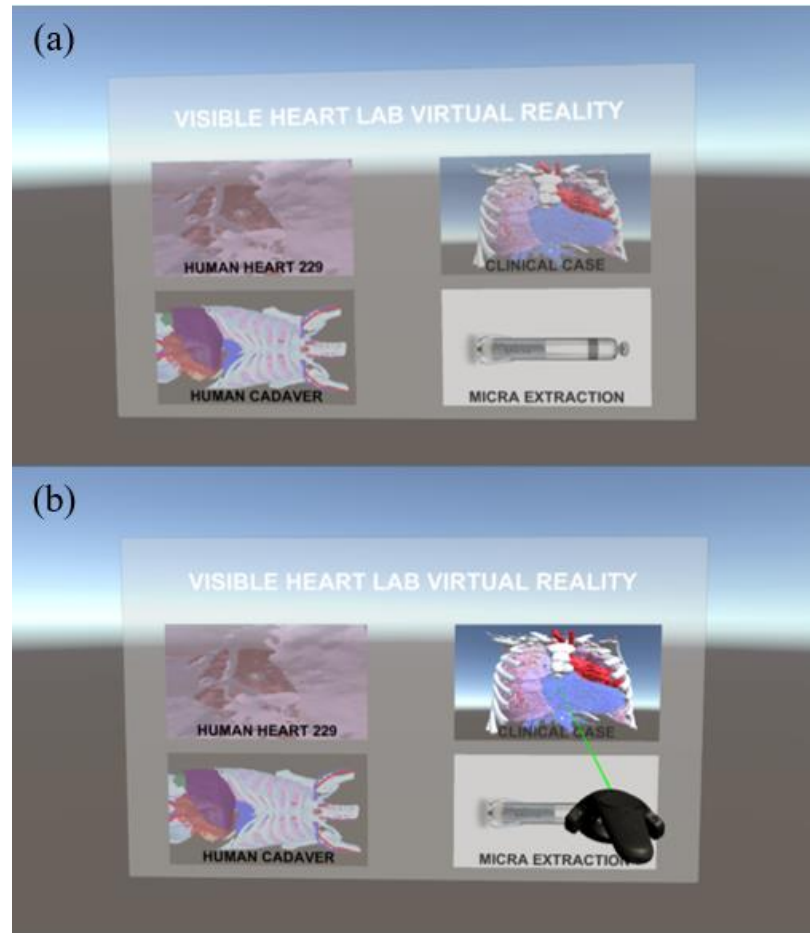

**Figure 4: (a) The menu displaying the scenes the user can enter. Depicted in (b) is the change in transparency of the button, highlighting a collision of the laser and the current button.**

A script and event handler manages the functionalities of users entering different virtual environments. An event handler is an empty game object that can be referenced from a user interface button in a unity scene. A script with custom methods was attached to the game object that specifies what happens when a button is selected from the menu (Figure 5). Clicking the back trigger on the Vive controller while highlighting a button calls a method in the event handler that starts the server (if one has not been created) or connects to the server (if a server has already been created). Once a server has been created, or connected to, the even handler loads the selected scene. Once within the scene, the user's transform information is replicated by their respective avatars to every client/connect user. A given user can seamlessly return to the main menu at any time by pressing the menu button on the Vive controller. This removes their avatar from the scene, but has no other effect on the remaining clients.

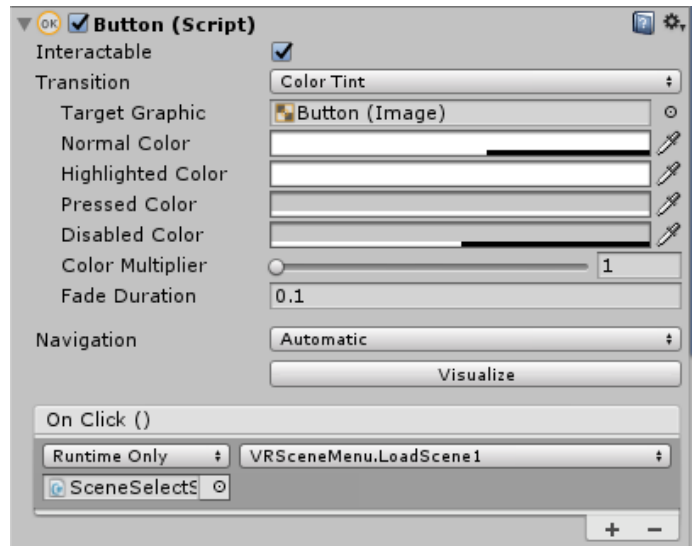

**Figure 5: Method for UI Button Scene Selection**

Finally, a custom script was written that allows the user to move throughout a given virtual reality environment. This script allows the user to move toward the laser on their controller by pressing down on the top half of the steam controller trackpad. Consequently, the user can also move backward by pressing the bottom half of the pad.

# Results

### *Anatomical Education*

The proposed system offers many benefits for medical and anatomical education or physician interactions. Work by others has already been done to study the effect of virtual reality on medical learning outcomes (Chang et al., 2017) with promising results. Currently, most students learn anatomy by reading textbooks and studying two-dimensional pictures. This is disadvantageous because you lose any three-dimensional spatial relationships of these complex anatomical features. The virtual reality system I am developing here within the Visible Heart® Laboratory, offers an accessible, immersive environments to learn complex human anatomies. Since the system user is in a three-dimensional environment, they can learn both varied anatomies and the spatial relationships of anatomical features simultaneously. The scale of the given anatomical models in the virtual scenes can be increased, even to a scale where the user can fly through them. This allows users to easily study the critical details of anatomical features.

As one system example, we created a virtual reality environment that contains a 3D model of a human heart (Figure 6). The scale of the human heart was increased so the user can easily fly through and explore the whole endocardial and epicardial features of the heart, granting the freedom to analyze features however they please. We believe and have obtained numerous user feedback from students, residents and experienced physicians that new insights can be gained by viewing these cardiac anatomies in this unique perspective. The laser pointer can highlight anatomical features to others in the scene, allowing for real-time instruction and collaborative discovery in an immersive virtual environment.

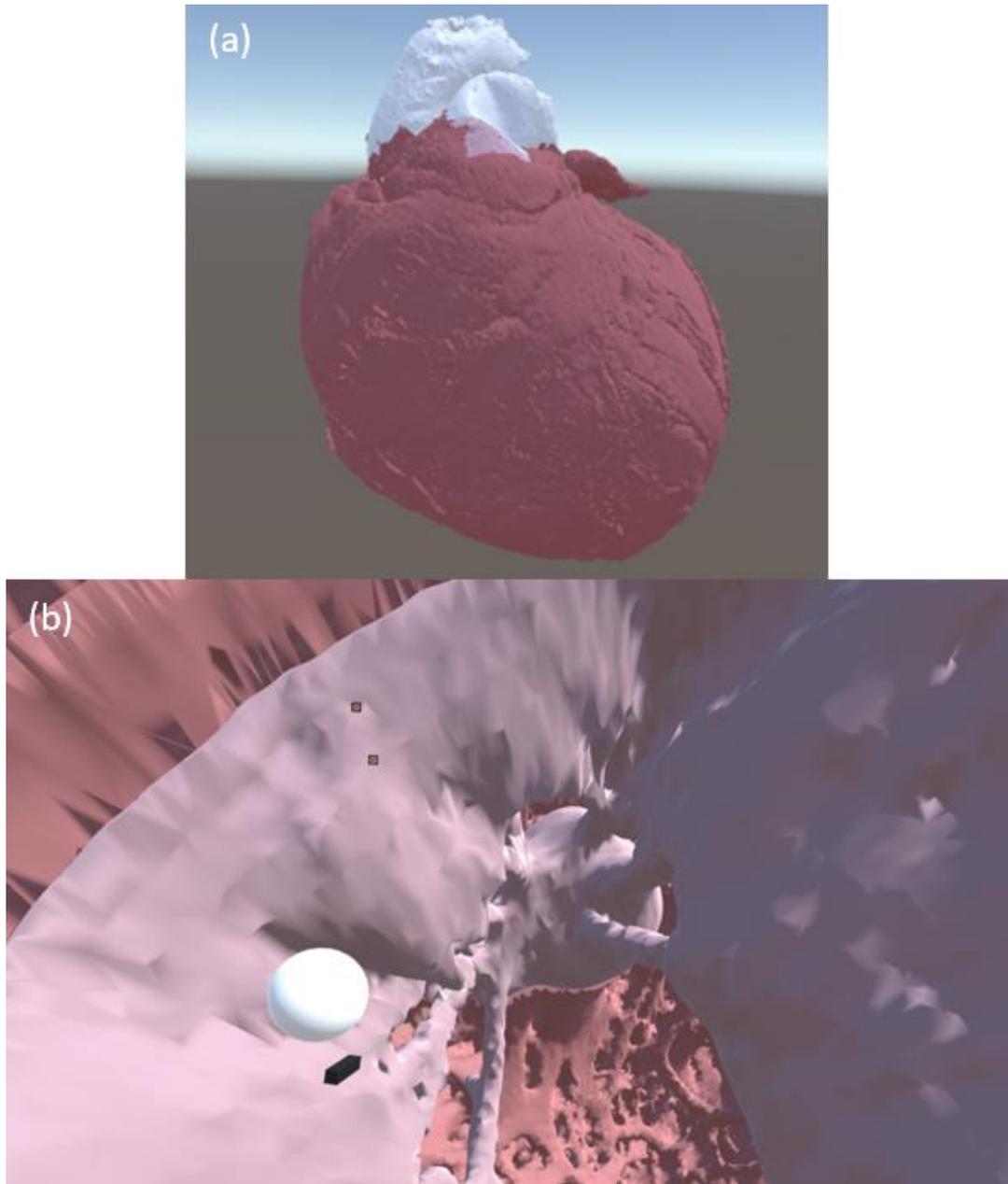

**Figure 6: (a) An external view of a computationally modeled human heart in the virtual reality environment. (b) Shown here, two users in the left atrium are studying the anatomy of the mitral valve. The instructor in the scene is using their**

**laser to highlight the large papillary muscle while explaining its function to the other users.**

We can also create scenes which are closer to surgical realism as well. For example, we have placed a model, created from a full human cadaver CT scan, into a virtual environment. This allows users to move around this cadaver at native scale as if it were lying on a surgical table (Figure 7a). This offers information on the actual spatial relationships between the many anatomical features modeled. Given the complex natures of human anatomy, allowing for a paired student-teacher interface can be highly instructive; implicating this setup as an education system as well. Further, scaling functionality is also advantageous in the teaching paradigm. For example, users can enlarge the size of the cadaver to the point where users can fly through the spinal canal (Figure 7b).

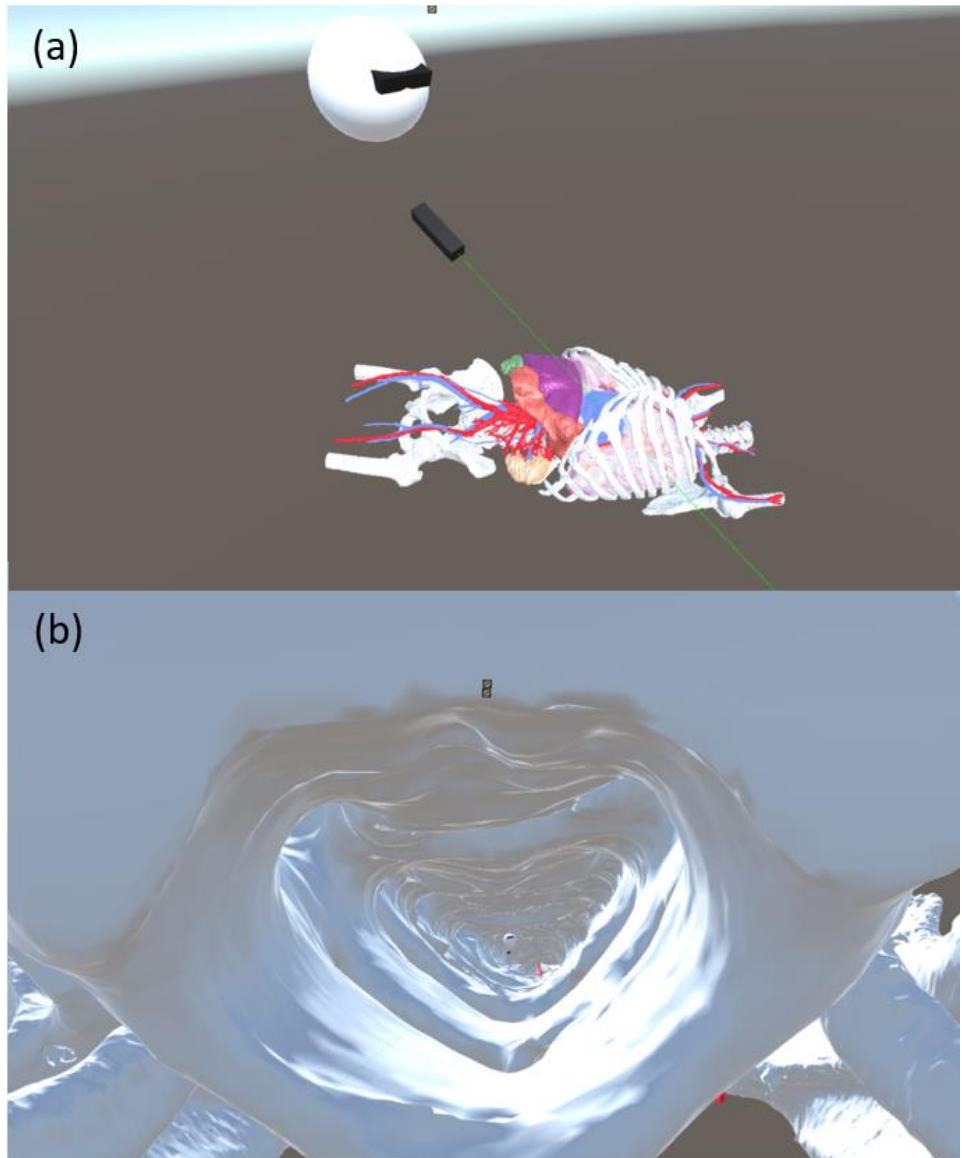

**Figure 7: (a) An example of an instructor teaching anatomy to others on a cadaver model, at native scale. (b) The scale of the cadaver has been enlarged and users can be seen navigating within the spinal canal.**

### *Pre-Surgical Planning*

For pre-surgical planning, currently the most relevant applications involve patients with relatively complex anatomy. For example, pediatric cardiac surgery related to congenital heart defects is one such example; due to both their reduced sizes and required complex repairs. To demonstrate this, we modeled a clinical scan from a pediatric patient with a large ventricular septal defect and inversion of the great vessels (Figure 9). This was created from pre-procedural clinical CT angiography, allowing for a clear image of the patient's blood volumes in the given heart chambers. From this perspective, it is easy to see the persisting patent ductus arteriosus as well as the ventricular septal defect. The virtual environment can be manipulated, allowing for certain objects to be shown or hidden. In this scene, the clinical care team can view the heart with the skeleton and lungs modeled. This offers an immersive look at the relationships between this patient's heart and other anatomical features. The rest of the abdomen can also be hidden, offering an unobstructed look at the heart anatomy and clinical features of the congenital defects. This model can also be arbitrarily scaled allowing subtle anatomical features to be more easily visualized.

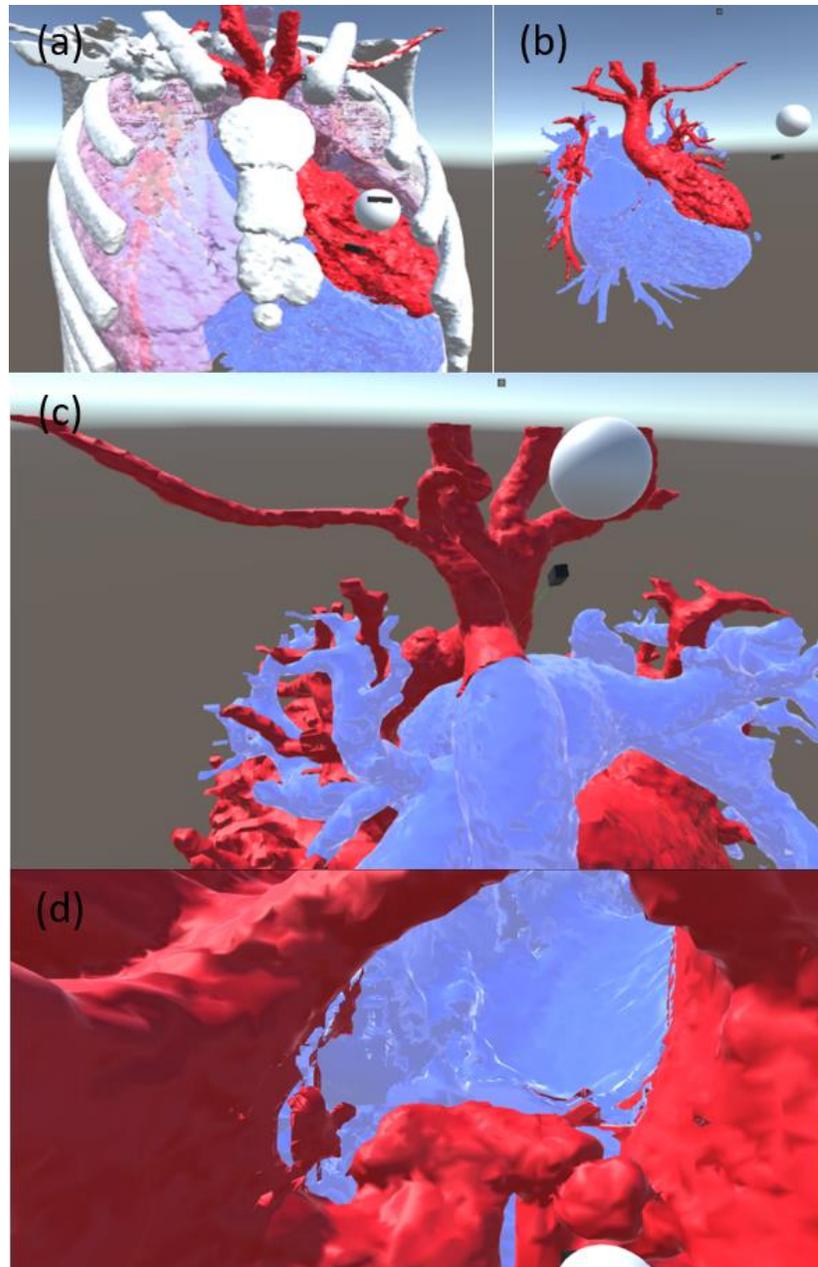

**Figure 8: (a) External view of a computationally modelled pediatric congenital heart disease case. (b) The lungs and skeleton have been removed to better depict the features of this congenital heart. (c) The clinical care team can analyze the inversion of the great vessels. (d) Collaborators standing in the right ventricle study the ventricular septal defect.**

***Medical Devices***

Frequently, understanding the detailed spatial relationships between medical devices and patient anatomies can be informative for both advancing medical device refinements and utilizations. This pipeline can use models obtained by imaging of anatomies with devices implanted or can also allow for devices to be computationally implanted in detailed anatomical models, and then either can be analyzed by collaborators in virtual environments. More specifically, medical device collaborators can enter a virtual environment to view their device interacting with the anatomy in these novel developed scenes. Biomedical Engineers can also utilize this collaborative methodology for proposing their device to physicians, either explaining the planned procedures in an immersive environment or to receive crucial feedback for improving their devices. To demonstrate this, we have compiled a 3D human heart model with a computationally added Medtronic Micra™ and extraction tool (Figure 10). This environment depicts a Micra™ extraction procedure (Vatterott et al., 2018). This modality yields a new perspective for studying this emerging procedure; i.e., the devices involved, and their interactions with a given patient's cardiac anatomy.

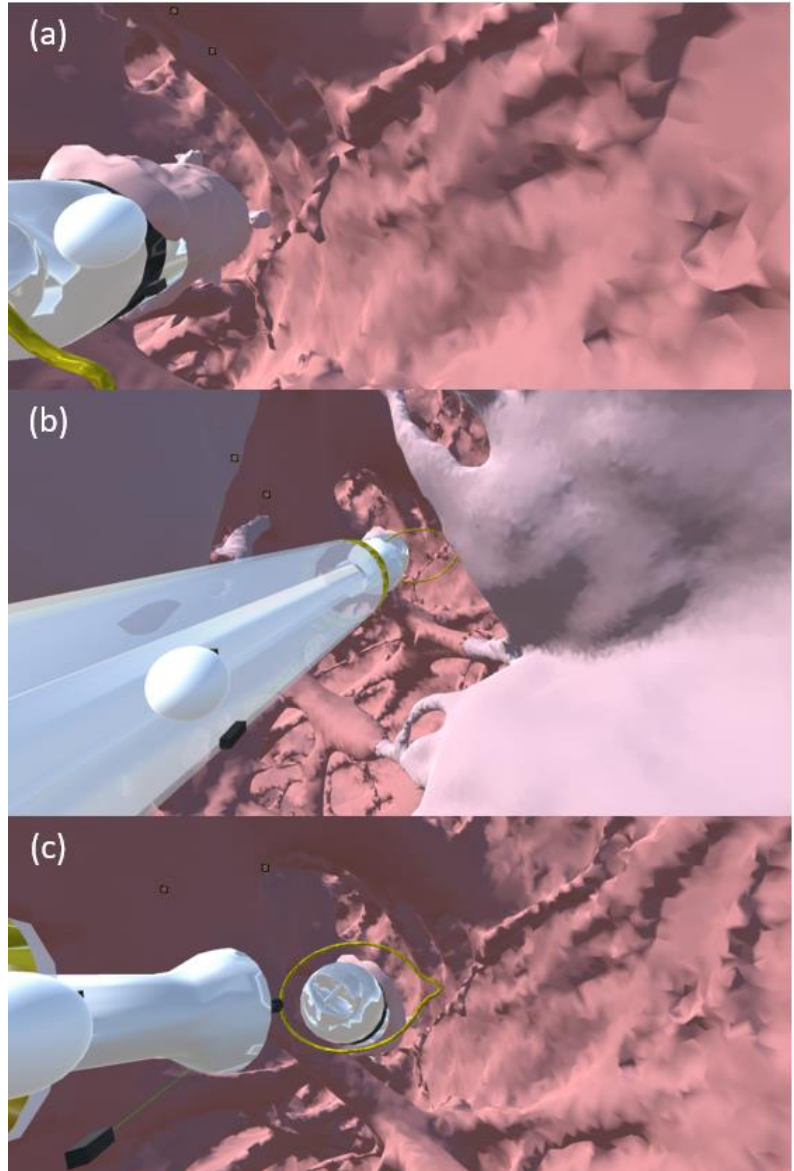

**Figure 9: (a) A collaborator (e.g., medical device developer) showing the encapsulation of the Micra™ tines in the right ventricular myocardium to other collaborators within this unique human heart environment. (b) Collaborators in the right atrium looking at the Micra™ and extraction tool through the tricuspid valve. (c) A user in the right ventricle highlighting the lasso on the distal end of the extraction tool that is employed to snare the Micra™.**

Maintaining acceptable network performances is critically important for allowing and maximizing these collaborative interfaces. Luckily, our developed pipeline only requires management of simple transform information, making data transfer overhead relatively minimal.

# Discussion

The ability to visualize, manipulate, and synchronize environment states with multiple participating users has applications for a large class of both educational and clinical problems. Broadly, it represents the novel abilities to distribute more salient information for faster human processing and communication. Pre-clinical planning and education represent a couple of important examples that are the most approachable given the current state of technology. However, there are technological limitations that prevent optimized utilizations of such a system. Segmentation is a time-consuming task, requiring the skills of highly trained anatomists, and characterized by uncertainty regarding human variance. Scaling of such a system would require greater optimization of this task through advances in computer vision or artificial intelligence. It is also unknown to what extent post-processing creates models which diverge from the morphologies of their native anatomies. More research into validation schemes will also be necessary to overcome this problem. Nevertheless, all generated scenes can be shared as educational tools for those who could benefit from their unique applications. Yet, if one can fill these dependencies, frameworks for clinical collaborations will be more feasibly scaled.

These applications are not limited to preclinical or anatomical educational tasks. The emergence of the field of surgical robotics, although in its infancy, has the difficult objective of creating systems which can localize and navigate human anatomy. Ultimately, this will require algorithms which can understand and act upon real three-dimensional spaces. However, gaining information about human anatomy is difficult without the availability of training data. From this perspective, our setup has utility for making progress for such in two main ways: it creates an interactive system for a computer agent to explore, and it creates a substrate for live 3D updates during a purported intra-surgical procedure. Although this has ultimate utility for automated technologies, more informative navigation is immediately useful for practicing physicians as well.

In the field of 3D anatomical modeling, it is frequently useful to take measurements of specific structures to inform the proper selection of a medical device to be employed. For example, selection of an artificial heart valve could be informed by the detailed analyses and 3D understanding of the dimensions and relative shapes of the patient's native annulus. Other specialized 3D measurement tools would be important in a variety of fields such as neurosurgery, orthopedic surgery, oral surgery, and more. Given the flexibility of tools present within the Unity3D or other video game engines, these technologies could easily be incorporated to increase utilities. Yet, additional discussions relative to such future work within each respective field is beyond the scope of this thesis.

To date, our laboratory has not attempted to share any information across the network other than transformation information. However, collaboration is highly dependent of vocal communication, which we do not provide. Yet, this can be solved simply by using a phone or voice chat app in tandem with our system. There are also solutions which allow voice data to be shared easily over the same network which was used for transforming data. We again leave refinements of this system to our group's future work.

While the field of virtual reality and augmented reality has progressed exponentially, their mutual utilizations by multiple uses has been minimal. Yet, recent projects have developed scenes which allow a Vive and HoloLens (Microsoft Corporation, Redmond, WA, USA) to occupy the same set of spatial coordinates. Such a setup has promise for taking advantage of the benefits of either or both virtual and augmented realities. This would allow for collaboration between modalities or even for a HoloLens user to manipulate objects using Vive controllers. This has already been implemented as a demo, but their utilizations in the medical field have so far absent. As new products emerge from this market, more possibilities, like this, will become available, expanding the repertoire for how visualization problems will be solved.

## Conclusion

Uses of virtual reality as a tool in telemedicine is a new and promising technique for the facilitation of physician collaboration and education over long distances. These technologies will permeate many aspects of medical inquiry, including live surgical cases, but there are many technological dependencies that must addressed before their uses becomes ubiquitous. This paper outlines just one necessary contribution so to expand these 3D environments. Fortunately, the breadth of applications for this technology is extensive enough to allow for substantial creativity. The hardware and software necessary for building these tools will also improve quickly and decrease in their costs, given the current attention given to the field, especially in the field of video game development. Consequently, it will be important to apply advances from external domains into the field of medicine.

Further work will now require customizations of scenes such as those presented here: i.e., through the development of specialized networking and manipulation tools. Currently, cardiac surgery, orthopedic surgery, neurosurgery, and oral surgery appear to be good candidates, however, these are still limited by lack of automation regarding segmentation and modeling of clinical anatomies. Nevertheless, our current pipeline provides a foundation that is immediately applicable to investigators with complex cases or researchers seeking novel techniques to visualize and interact with anatomical structures.

# Appendix

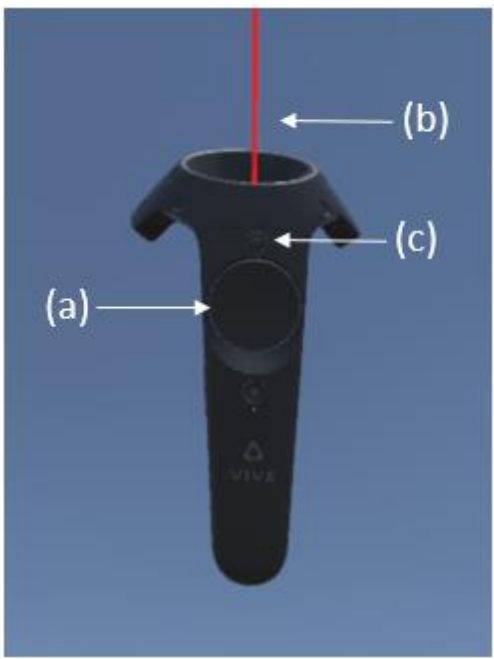

**Appendix Figure 1: An image of the HTC Vive Controller depicting (a) the trackpad, (b) the laser, and (c) the menu button.**